\documentclass[preprint]{aastex}

\begin{document}

\title {The Pre-He White Dwarfs in Eclipsing Binaries. I. WASP 0131+28 }
\author{Jae Woo Lee$^{1}$, Jae-Rim Koo$^{2}$, Kyeongsoo Hong$^{3}$, and Jang-Ho Park$^{1}$ }
\affil{$^1$Korea Astronomy and Space Science Institute, Daejeon 34055, Republic of Korea}
\affil{$^2$Department of Astronomy and Space Science, Chungnam National University, Daejeon 34134, Republic of Korea}
\affil{$^3$Institute for Astrophysics, Chungbuk National University, Cheongju 28644, Republic of Korea}
\email{jwlee@kasi.re.kr}

\begin{abstract}
We report the first $BV$ light curves and high-resolution spectra of the post-mass transfer binary star WASP 0131+28 to study 
the absolute properties of extremely low-mass white dwarfs. From the observed spectra, the double-lined radial velocities were 
derived, and the effective temperature and rotational velocity of the brighter, more massive primary were found to be $T_{\rm eff,1} 
= 10,000 \pm 200$ K and $v_1\sin$$i$ = 55 $\pm$ 10 km s$^{-1}$, respectively. The combined analysis of the {\it TESS} archive data and 
ours yielded the accurate fundamental parameters of the program target. The masses were derived to about 1.0 \% accuracy and the radii 
to 0.6 \%, or better. The secondary component's parameters of $M_2 = 0.200 \pm 0.002$ M$_\odot$, $R_2 = 0.528 \pm 0.003$ R$_\odot$, 
$T_{\rm eff,2}$ = 11,186 $\pm$ 235 K, and $L_2 = 3.9 \pm 0.3$ L$_\odot$ are in excellent agreement with the evolutionary sequence for 
a helium-core white dwarf of mass 0.203 M$_\odot$, and indicates that this star is halfway through the constant luminosity phase. 
The results presented in this article demonstrate that WASP 0131+28 is an EL CVn eclipsing binary in a thin disk, which is formed 
from the stable Roche-lobe overflow channel and composed of a main-sequence dwarf with a spectral type A0 and a pre-He white dwarf. 
\end{abstract}


\section{INTRODUCTION}

Over 97 \% of all the stars in the Milky Way, including our Sun, are thought to end their evolution as white dwarfs (WDs) 
(Fontaine et al. 2001; Woosley \& Heger 2015). Because the WD stars are abundant and long-lived, they provide a powerful technique for 
studying the stellar age and formation history of various Galactic populations (Temmink et al. 2019). Most of them are carbon-oxygen (CO) 
core and hydrogen-rich envelope WDs with a mass distribution clustered around $\sim$0.6 $M_\odot$ (Kepler et al. 2007). The CO-core WDs 
are the most likely outcome of single-star evolution. However, extremely low-mass (ELM) WDs with masses below 0.3 $M_\odot$ cannot be 
formed from isolated stars, because for a single star their progenitors would require evolutionary time longer than the current age of 
the universe. The best explanation for such ELMs is the binary evolution channel (Marsh et al. 1995; Kilic et al. 2007). Generally, 
ELM WDs are produced after significant mass loss during the red giant branch stage before helium (He) burning, through either 
the stable Roche-lobe overflow (RL channel) or the unstable common-envelope ejection (CE channel) in binary stars (Althaus et al. 2013; 
Chen et al. 2017, Li et al. 2019). As a consequence of the non-conservative mass transfer, the donor stars become stellar remnants with 
insufficient mass to ignite the He in their cores, and are known as He-core WDs.  

EL CVn-type stars are a special group of post-mass transfer eclipsing binaries (EBs) first found by Maxted et al. (2011, 2014a). 
They are composed of an A/F-type main sequence (MS) dwarf and an ELM WD precursor (hereafter pre-He WD). The latter is also 
called a stripped red giant, proto-He WD, or thermally bloated WD. About 70 EBs have been reported to be EL CVn-like stars 
with binary orbital periods ($P$) from 0.46 to 5.2 days (Maxted et al. 2014a; van Roestel et al. 2018; Wang et al. 2020), 
except for KOI-81 ($P$ = 23.8776 days, van Kerkwijk et al. 2010). In most cases, their light curves display boxy primary eclipses 
and ellipsoidal modulations outside eclipse caused by gravitational distortion of the companion star. Some pre-He WDs and their companions 
exhibit multiperiodic oscillations (Maxted et al. 2013, 2014b; Zhang et al. 2016), which are very useful for exploring the interior 
structure of these pulsators through asteroseismology. The evolution study of Chen et al. (2017) showed that EL CVn EBs are produced 
by the long-term RL channel. The CE channel causes stellar merging of binary components into a single star. Lee et al. (2018) suggested 
that R CMa-type EBs (Budding \& Butland 2011; Lee et al. 2016) are the precursors of the EL CVn binaries. By stopping mass transfer and 
contracting, semi-detached R CMa systems become detached ones, such as KIC 10661783 (Lehmann et al. 2013), KIC 8262223 (Guo et al. 2017), 
OO Dra (Lee et al. 2018), and KIC 7368103 and KIC 8823397 (Wang et al. 2019). Then, the detached R CMa binaries can be regarded as 
newly-born EL CVn binaries evolving into He WDs at about constant luminosity. At present, ten EBs have been identified as 
the R CMa types (Wang et al. 2019), eight of which are pulsating EBs. KIC 6206751 contains a $\gamma$ Dor star (Lee \& Park 2018) 
and the other systems contain a $\delta$ Sct variable. 

The evolutionary process of the ELM WDs may be quite different from that of more massive WDs. Both EL CVn and R CMa binaries are 
very promising targets for understanding the physical properties of the He-core WDs. However, the pre-He WDs are not easy to 
detect in spectroscopic observations because they are much fainter than their MS companions. Prior to this study, only eleven systems 
(3 in EL CVn and 8 in R CMa) had credible physical parameters calculated from double-lined radial velocities (RVs). In order 
to investigate the pre-He WDs in EBs, we have been performing their high-resolution spectroscopy and/or photometry. Here, we present 
the EL CVn-type star 1SWASP J013129.76+280336.5 (WASP 0131+28) as our first result, which is one of 17 EL CVn candidates identified 
by Maxted et al. (2014a) using the WASP photometric survey. From model fits to the WASP light curve and the observed flux distribution, 
they reported that the binary system had an eclipsing period of $P$ = 1.882752 days, an inclination angle of $i$ = 86.5 deg, 
a mass ratio of $q$ = 0.075, relative radii of $r_1$ = 0.2100 and $r_2$ = 0.0610, and effective temperatures of $T_1$ = 9500 K and 
$T_2$ = 10,500 K. The subscript 1 indicates the larger but cooler primary star (WASP 0131+28 A), while the companion is identified 
by subscript 2 (WASP 0131+28 B). 

The main goal of the research is to present the absolute properties of WASP 0131+28 (TIC 18225199; TYC 1755-509-1; $T_{\rm p}$ 
= $+$10.917; $V\rm_T$ = $+$10.97, $(B-V)\rm_T$ = $+$0.12) from detailed analyses of our time-series photometry/spectroscopy and 
the {\it TESS} public data, and to investigate its possible evolutionary state. We organize this paper as follows. 
The observations and spectral analysis are given in Sections 2 and 3, respectively. Section 4 presents the light-velocity solutions 
and system parameters. We end with discussion and conclusions in Section 5.

\section{OBSERVATIONS AND DATA REDUCTIONS}

\subsection{{\it TESS} PHOTOMETRY}

{\it TESS} (Ricker et al. 2015) observed our target star WASP 0131+28 in camera 1 during Sector 17. The photometric observations 
were obtained at 2 minutes cadence between 2019 October 8 and November 2 (BJD 2,458,764.67 $-$ 2,458,789.70). The simple aperture 
photometry (\texttt{SAP$_-$FLUX}) data from MAST\footnote{https://archive.stsci.edu/} were detrended following the procedure 
of Lee et al. (2017), and these fluxes were changed to {\it TESS} magnitudes by applying an apparent magnitude of 
$T_{\rm p}$ = $+$10.917 (Stassun et al. 2018, 2019). In this work, we used 12,793 individual measurements after removing 
106 data points as outliers, which were filtered with 5-$\sigma$ criterion and selected by visual inspection (Lee et al. 2019). 
The {\it TESS} light curve of WASP 0131+28 is displayed as black circles in Figure 1.

\subsection{GROUND-BASED PHOTOMETRY AND SPECTROSCOPY}

We carried out multi-band photometric and high-resolution spectroscopic follow-up observations for WASP 0131+28. The photometric 
$BV$ images were taken from 2016 to 2017 with an ARC 4K CCD camera on the 1.0-m telescope at Mt. Lemmon Optical Astronomy 
Observatory (LOAO; Lee et al. 2012) in Arizona, USA. We applied bias, dark, and flat-fielding corrections to the CCD frames using 
the IRAF/CCDRED package, and performed aperture photometry with the Python package PHOTUTILS\footnote{http://photutils.readthedocs.io/} 
(Bradley et al. 2017). Because no standard star field was taken for photometric calibration, we standardized seven stars in 
our observed images using the APASS DR9 catalogue (Henden et al. 2016). The standard magnitudes of our time-series data were 
calibrated by applying the ensemble normalization method of Gilliland \& Brown (1988). From the LOAO photometry, we obtained 
12,708 individual observations (6372 in $B$ and 6336 in $V$), which are provided in Table 1 and shown as gray circles in Figure 1. 
Table 2 presents the standard magnitudes of $V$ and the color indices of $B-V$ at four characteristic phases. 

Fifty high-resolution spectra were secured from 2016 to 2019 with the 1.8-m telescope and fiber-fed echelle spectrograph BOES at 
Bohyunsan Optical Astronomy Observatory (BOAO; Kim et al. 2007) in Korea. Because WASP 0131+28 is not bright enough to acquire 
high signal-to-noise ratio (SNR) spectra, we employed the $2\times2$ binning mode and the largest optical fiber of 300\,$\micron$ 
which results in a resolution of $R$ = 30,000 in a field of view of 4.3\arcsec. The exposure times of our program 
target was mostly set to be 40 min, corresponding to $\sim$0.015 in the orbital phase. ThAr arc and Tungsten-Halogen lamp 
(THL) images were taken every night for the identification of spectral lines and the blaze function calibration. 
After bias subtraction, we used the LACos routine written by van Dokkum (2001) to remove cosmic ray hits. Aperture extraction and 
wavelength calibration were conducted using the IRAF/ECHELLE package. Almost all spectra have an SNR larger than 40 (on average 55) 
at around 5500 {\AA}.

\section{SPECTRAL ANALYSIS} 

The BOES spectra were used to determine the radial velocities (RVs) of WASP 0131+28. Figure 2 presents the trailed spectra of 
the binary system in the \ion{Mg}{2} $\lambda$4481 region, where the absorption lines from WASP 0131+28 A and WASP 0131+28 B are 
clearly identified and shifted along the orbital phases. We applied the two dimensional cross-correlation (TODCOR) algorithm 
(Zucker \& Mazeh 1994) to simultaneously measure the RVs of both components. The synthetic spectra matching each star were derived 
from the Kurucz models (Castelli \& Kurucz 2003) by adopting the stellar parameters of Maxted et al. (2014a) at the beginning and 
ours at the end. The two spectra were shifted as a RV function of each component and were convolved considering the light ratio of 
$L_1/(L_1+L_2) \simeq 0.9$. We obtained the RVs minimizing the difference between the convolved and observed spectra. 
The resulting RV measurements and their errors are given in Table 3 and displayed in Figure 3. 

To get the atmospheric parameters of WASP 0131+28, it is necessary to educe each component's spectrum from the observed spectra. With 
the FDB\textsc{inary} code (Iliji\'c et al. 2004)\footnote{\url{http://sail.zpf.fer.hr/fdbinary}}, we obtained a disentangling spectrum 
with a better SNR for the primary component and performed $\chi^{2}$ minimization between the reconstructed and model spectra. 
This procedure is very similar to that employed by Hong et al. (2017). For measuring the temperature ($T_{\rm eff,1}$) and 
the rotational velocity ($v_1\sin$$i$), we chose eight absorption lines (\ion{Ca}{2} K $\lambda$3933, H$_{\rm \delta}$, 
\ion{Ca}{1} $\lambda$4226, \ion{Fe}{1} $\lambda$4271, H$_{\rm \gamma}$, \ion{Fe}{1} $\lambda$4383, \ion{Mg}{2} $\lambda$4481, 
and H$_{\rm \beta}$). A total of 44,000 synthetic spectra were generated in ranges of 8000 $\le$ $T_{\rm eff,1}$ $\le$ 12,000 K and 
10 $\le$ $v_1\sin$$i$ $\le$ 120 km s$^{-1}$ from the BOSZ spectral library\footnote{\url{https://archive.stsci.edu/prepds/bosz}} 
(Bohlin et al. 2017). For the model spectra, solar metallicity was assumed, and the microturbulent velocity and surface gravity were 
assumed to be 2.0 km s$^{-1}$ and log $g_1$ = 4.2 (cf. Section 4), respectively. The consequential $\chi^{2}$ diagrams are plotted 
in Figure 4, and the best-fitting values of $T_{\rm eff,1}$ and $v_1\sin$$i$ were found to be $10,000\pm200$ K and $55\pm10$ km s$^{-1}$, 
respectively. The errors on these parameters were estimated from the 1$\sigma$-values of the optimal values in the eight regions. 
The primary's temperature is in satisfactory agreement with the value ($9500\pm700$ K) estimated by Maxted et al. (2014a). 
Figure 5 presents the eight absorption regions of the reconstructed spectrum, where there are overlaid our optimal model spectrum. 
On the other hand, the atmospheric parameters of the faint secondary cannot be estimated because its disentangled spectrum has 
very low SNR.

\section{BINARY MODELING AND ABSOLUTE DIMENSIONS}

In double-lined EBs, masses and radii can be measured at the few percent level or better from detailed analyses of light and 
RV curves. As a result, these quantities can be used to understand the absolute properties of the stars and to improve their structure 
and evolution models (Hilditch 2001; Torres et al. 2010). Maxted et al. (2014a) analyzed the WASP light curve of WASP 0131+28 using 
the JKTEBOP code (Southworth 2010) and classified it as an EL CVn star with characteristics of both short $P$ and low $q$. 
As presented in Figure 1, the LOAO and {\it TESS} light curves also are typical of the EL CVn type. For the light and velocity solutions 
of WASP 0131+28, we adopted the 2007 version of the Wilson-Devinney program (Wilson \& Devinney 1971, van Hamme \& Wilson 2007; W-D07). 
Our $BV$ and RV curves were modeled with the {\it TESS} public data in almost the same way as the R CMa-type binary OO Dra 
(Lee et al. 2018). In this synthesis, we simultaneously solved all individual observations using the detached mode 2 of this program. 

The effective temperature of WASP 0131+28 A was initialized at $T_1$ = 10,000 K from an analysis of our disentangling spectrum. 
The bolometric albedos $A_{1,2}$ and gravity-darkening exponents $g_{1,2}$ were all set to be 1.0, which is suitable for 
components with radiative atmospheres. The logarithmic bolometric ($X_{\rm bol}$, $Y_{\rm bol}$) and monochromatic ($x_{\rm passband}$, 
$y_{\rm passband}$) limb-darkening coefficients were taken by interpolation from the tables of van Hamme (1993). 
This synthesis was iterated until the correction to each free parameter is lower than the corresponding standard deviation ($\sigma$) 
calculated by the W-D07 program. The light and RV solutions from this modeling are presented in columns (2)$-$(3) of Table 4, where 
the parameters with errors are the fitted parameters. To obtain the parameter errors, we split the LOAO and {\it TESS} time-series 
data into five subsets and separately analyzed them with the W-D07 program. Then, the 1$\sigma$-value of each free parameter was 
computed from the five different values and adopted as its error (Koo et al. 2014). The synthetic light and RV curves appear 
as red lines in Figure 1 and Figure 3, respectively, where the corresponding residuals are plotted in the lower panel of each figure. 
On the other hand, we tried to find any pulsation frequencies in the light residuals from the binary modeling, but no significant signal 
was detected. 

The absolute dimensions of WASP 0131+28 were directly computed without any assumptions from a combined solution of the three datasets 
({\it TESS}, LOAO, and BOES) collected in this work. They are presented in Table 5, where $R$ is the radii of spherical stars of 
the same volume tabulated by Mochnacki (1984). Bolometric magnitudes ($M_{\rm bol}$) and luminosities ($L$) were obtained by employing 
the solar values of $M_{\rm bol}$$_\odot$ = +4.73 and $T_{\rm eff}$$_\odot$ = 5780 K. Bolometric corrections (BCs) were taken from 
the empirical relations of Torres (2010) between $\log T_{\rm eff}$ and BC. We derived the distance of WASP 0131+28 to be 766 $\pm$ 35 pc, 
using the maximum light of $V$ = +10.82 $\pm$ 0.02 listed in Table 2 and the Galactic extinction of $A_{\rm V}$ = 0.214 from 
the reddening map of Schlafly \& Finkbeiner (2011). This distance is a good match to 741$\pm$33 pc, inverted from the $Gaia$ parallax 
of 1.349 $\pm$ 0.060 mas (Gaia Collaboration et al. 2018).

\section{DISCUSSION AND CONCLUSIONS}

We have presented follow-up photometry and spectroscopy of WASP 0131+28 and analyzed the data in detail, together with the {\it TESS } 
public data. The BOES spectra yielded an effective temperature of $T_{\rm eff,1} = 10,000 \pm 200$ K and a rotational velocity of 
$v_1\sin$$i$ = 55 $\pm$ 10 km s$^{-1}$. The synchronous rotation of $v_{\rm 1,sync}$ = 49.0 km s$^{-1}$ is within the $v_1\sin$$i$ value, 
which implies that the primary star is synchronized with the orbital period. Double-lined RVs were derived using the TODCOR algorithm 
of Zucker \& Mazeh (1994) using two synthetic spectra for each component. The binary star modeling indicates that WASP 0131+28 is 
a detached EB with parameters of $i$ = 85.027 deg, $q$ = 0.1005, and $T_2$ = 11,186 K. The component stars have the fill-out factors 
of $f_1$ = 42 \% and $f_2$ = 65 \%, which are defined as $f_{1,2}$ = $\Omega_{\rm in}$/$\Omega_{1,2}$. In the preceding section, 
we determined the fundamental parameters of our target star from the simultaneous analysis of all time-series data, as follows: 
$M_1 = 1.986 \pm 0.017$ M$_\odot$, $M_2 = 0.200 \pm 0.002$ M$_\odot$, $R_1 = 1.824 \pm 0.006$ R$_\odot$, $R_2 = 0.528 \pm 0.003$ R$_\odot$, 
$L_1 = 30 \pm 2$ L$_\odot$, and $L_2 = 3.9 \pm 0.3$ L$_\odot$. Along with the shape of the light curve, the low $q$ and 
$M_2 <$ 0.3 $M_\odot$ indicate that WASP 0131+28 is a typical EL CVn-type star. 

For the Galactic kinematics of WASP 0131+28, we determined the space coordinates ($X, Y, Z$) and velocity components ($U, V, W$) from 
its position ($\alpha$, $\delta$), proper motion, system velocity, and distance (Johnson \& Soderblom 1987). For this calculation, 
we adopted the position of ($X_\odot$, $Y_\odot$, $Z_\odot$) = ($-$8.0, 0.0, 0.0) for the Sun (Reid 1993), its motion of 
($U_\odot$, $V_\odot$, $W_\odot$) = (9.58, 10.52, 7.01) km s$^{-1}$ relative to the local standard of rest (LSR; Tian et al. 2015), and 
the LSR velocity of 220 km s$^{-1}$. The proper motion of our target star was taken to be $\mu _\alpha \cos \delta$ = 0.302 $\pm$ 0.092 mas yr$^{-1}$ 
and $\mu _ \delta$ = $-$8.197 $\pm$ 0.111 mas yr$^{-1}$ from the $Gaia$ DR2. Further, we used the ORBIT6 code of Odenkirchen \& Brosche (1992) 
to compute the angular momentum in the $z$ direction ($J_{\rm z}$) and the eccentricity ($e$) of the binary's Galactic orbit. The results 
are presented in Table 6. Pauli et al. (2006) classifies the population membership of WDs in the $U-V$ and $J_{\rm z}-e$ diagrams. 
In Figure 6, we clearly see that WASP 0131+28 belongs to a thin-disk population.

The accurate absolute parameters allow us to simulate the evolutionary history of stars. The locations of WASP 0131+28 AB in 
the Hertzsprung-Russell (HR) and $\log T_{\rm eff}-\log g$ diagrams are displayed in Figure 7, together with those of eight double-lined 
R CMa (Wang et al. 2019) and three EL CVn (WASP 0247-25, Maxted et al. 2013; KOI-81, Matson et al. 2015; EL CVn, Wang et al. 2020) EBs. 
The up- and down-pointing triangles represent semi-detached and detached R CMa-type stars, respectively, and the circles denote 
the EL CVn stars. In the same figure, we plotted the four evolutionary tracks of ELM WDs with masses of 0.179 M$_\odot$, 0.195 M$_\odot$, 
and 0.234 M$_\odot$ from Driebe et al. (1998) and a mass of 0.203 M$_\odot$ from Althaus et al. (2013). 
With the more massive primary components of the other EBs, WASP 0131+28 A resides inside the MS band between ZAMS and TAMS, and 
its companion of mass 0.200 $M_\odot$ is well-matched with the He-core WD model of 0.203 M$_\odot$. This indicates that WASP 0131+28 B 
is in the so-called constant luminosity phase, evolving to higher temperature and smaller radius at nearly constant luminosity. 
As illustrated in this figure, the semi-detached Algols with a short $P$, low $q$ and $M_2$ combination will eventually become 
EL CVn-type stars via the detached R CMa stage which is the starting place of the constant luminosity phase, as mass transfer terminates 
and the low-mass companions shrink by hydrogen shell burning. As mentioned in Section 1, Chen et al. (2017) suggested that the EL CVn EBs 
are formed through the stable RL channel but not the CE channel. In that case, a strong correlation exists between the orbital periods 
$P$ and the ELM WD masses $M_{\rm WD}$ (Lin et al. 2011). WASP 0131+28 B is a good match with the $P-M_{\rm WD}$ relation for the stable 
mass transfer episode given by Lin et al. (2011). This agreement supports the stable RL hypothesis for the EL CVn-type binaries.

\acknowledgments{ }
The authors wish to thank the staffs of BOAO and LOAO for assistance during our observations and Dr. Andreas Irrgang for providing 
a copy of the ORBIT6 code. This paper includes data collected by the {\it TESS} mission, which were obtained from MAST. Funding for 
the {\it TESS} mission is provided by the NASA Explorer Program. This research has made use of the Simbad database maintained at CDS, 
Strasbourg, France, and was supported by the KASI grant 2020-1-830-08. The work by K.H. and J.-R.K. was supported by the grant Nos. 
2017R1A4A1015178 and 2017R1A6A3A01002871, and 2017R1A6A3A01002871, respectively, of the National Research Foundation (NRF) of Korea.

\clearpage
\begin{figure}
\includegraphics{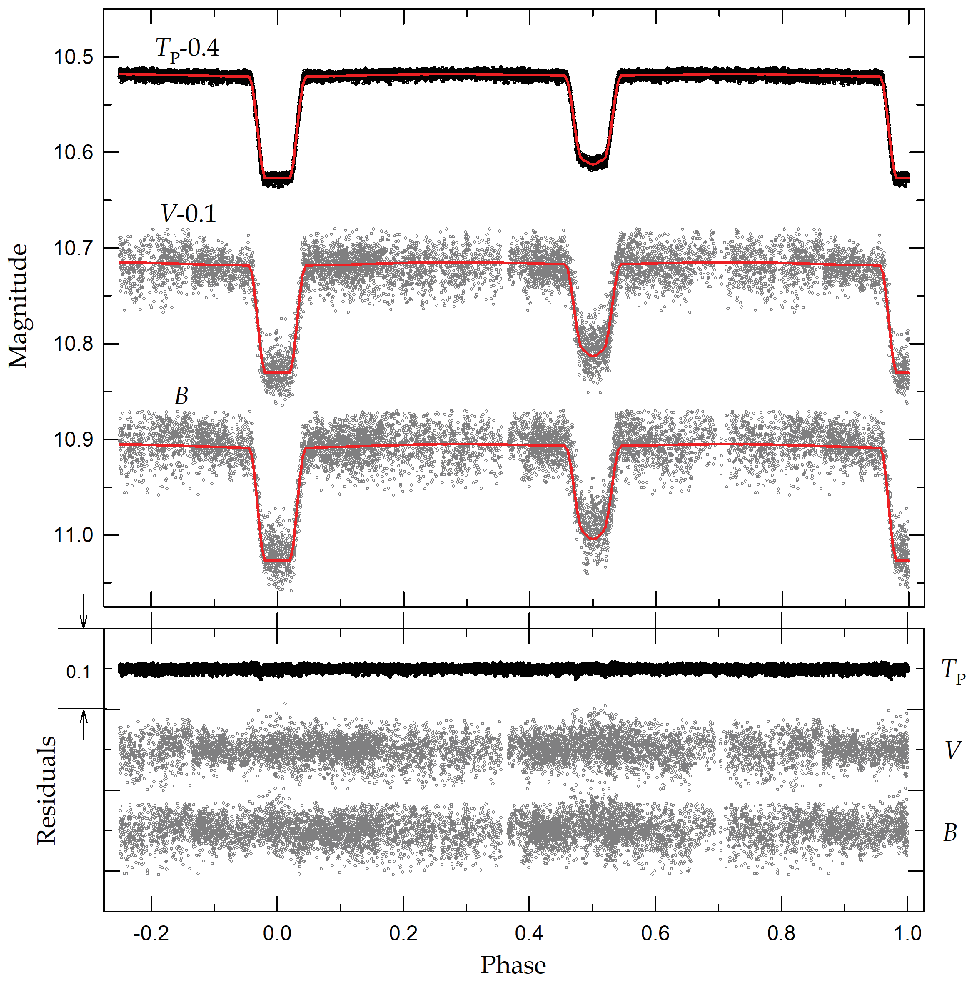}
\caption{Light curves of WASP 0131+28 with the fitted models. The black and gray circles are individual observations taken from 
{\it TESS} and LOAO, respectively, and the solid lines represent the synthetic curves obtained from a simultaneous analysis of 
all curves. The lower panel shows the differences between observations and models.}
\label{Fig1}
\end{figure}

\begin{figure}
\includegraphics[]{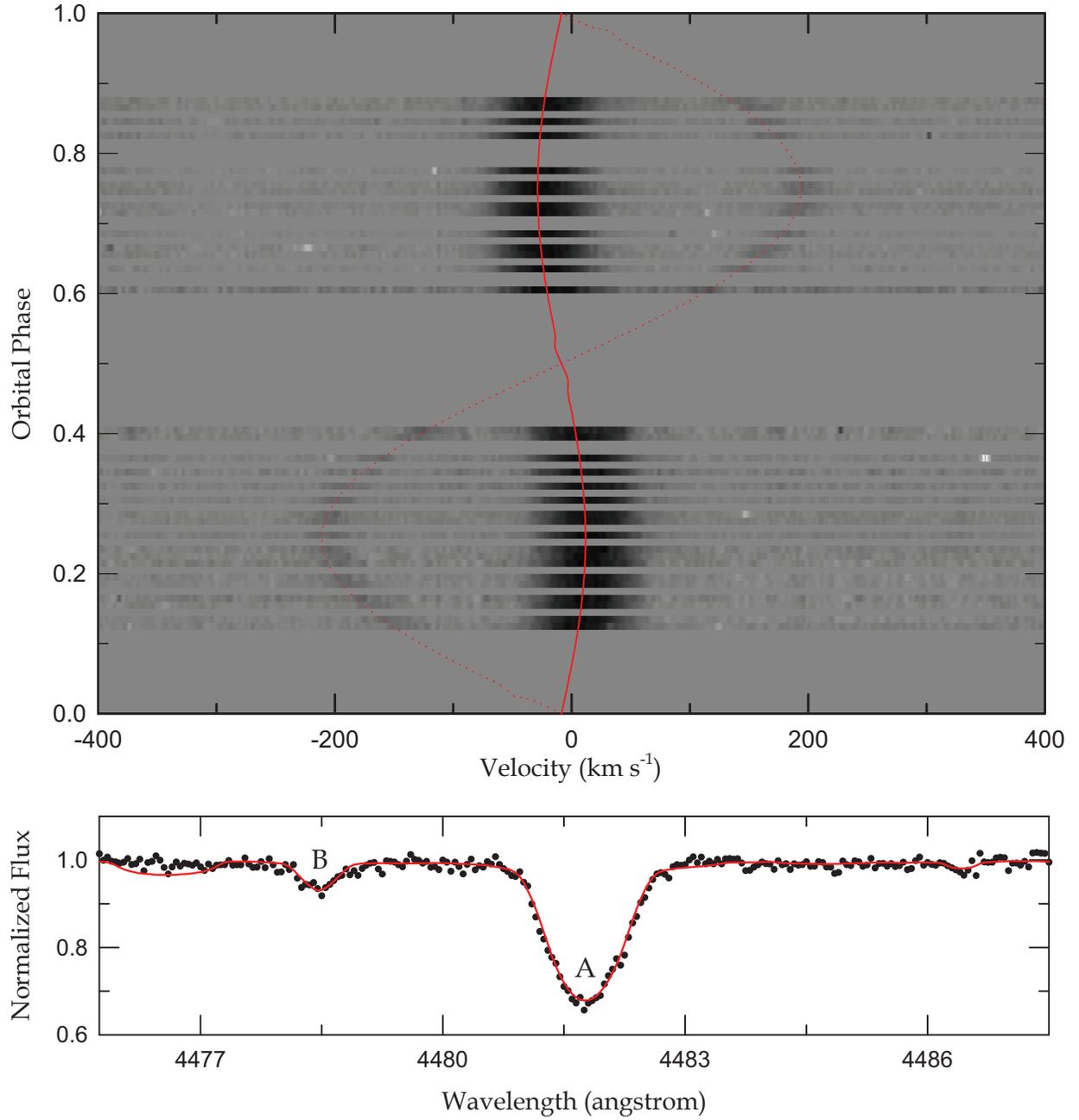}
\caption{Upper panel displays the trailed spectra of WASP 0131+28 in the Mg II region. The solid and dashed lines track 
the orbital motions of the primary (A) and secondary (B) components, respectively. In the lower panel, the circle and line 
represent the observed spectrum and its model fit at an orbital phase of 0.24 (BJD 2,457,809.9366), respectively. }
\label{Fig2}
\end{figure}

\begin{figure}
\includegraphics[]{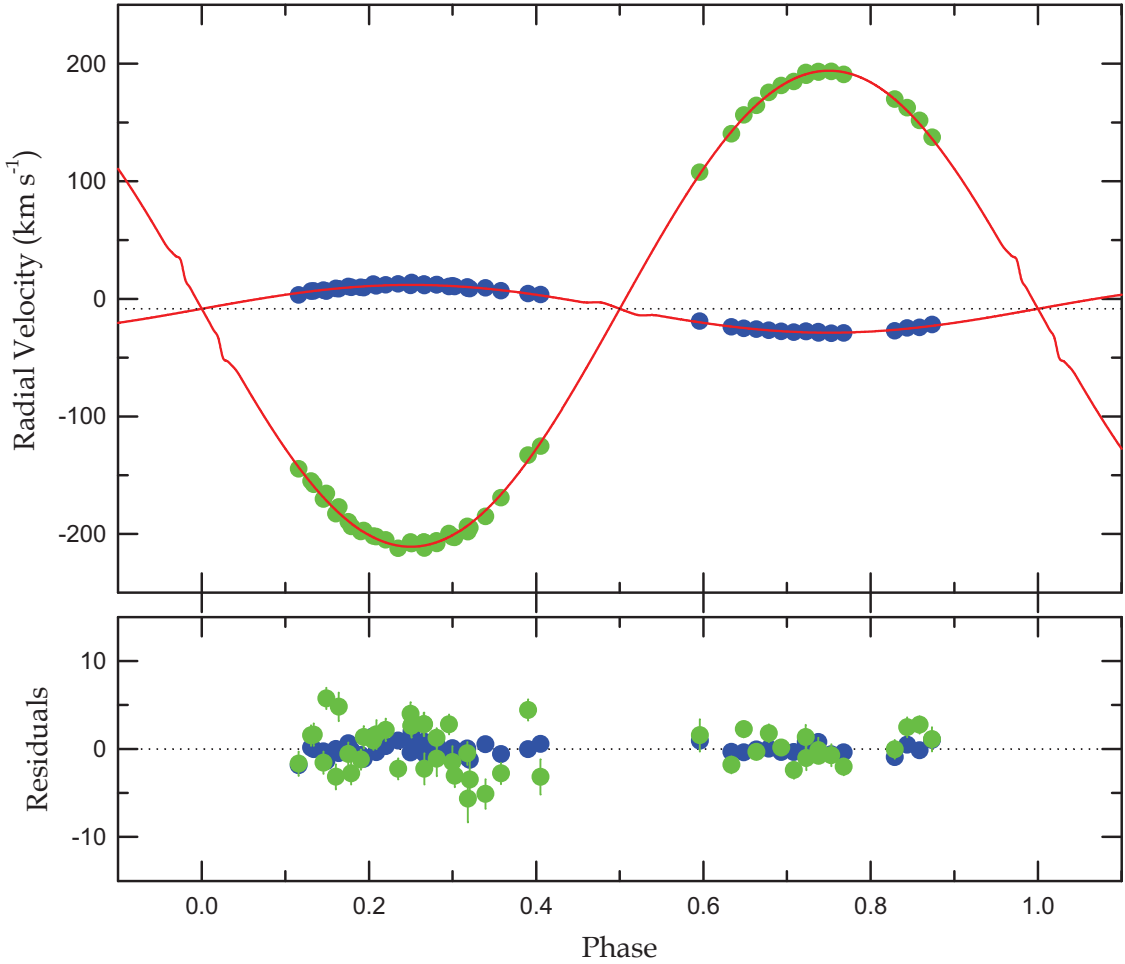}
\caption{RV curves of WASP 0131+28 with fitted models. The blue and green circles are the primary and secondary measurements, 
respectively, and the solid curves denote the results from a consistent light and RV curve analysis with the W-D07 program. 
The dotted line represents the system velocity of $-$8.44 km s$^{-1}$. The lower panel displays the residuals between measurements 
and models. }
\label{Fig3}
\end{figure}

\begin{figure}
\includegraphics[scale=0.9]{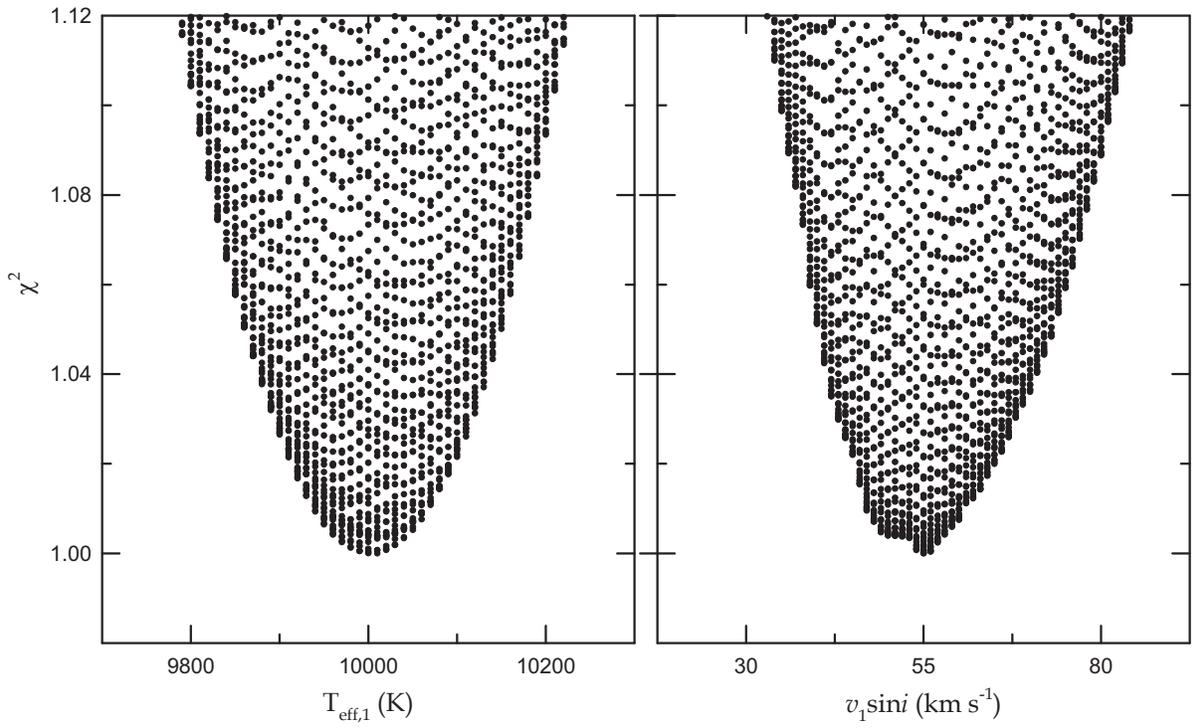}
\caption{$\chi^2$ diagrams of the effective temperature (left) and the projected rotational velocity (right) of the primary star. }
\label{Fig4}
\end{figure}

\begin{figure}
\includegraphics{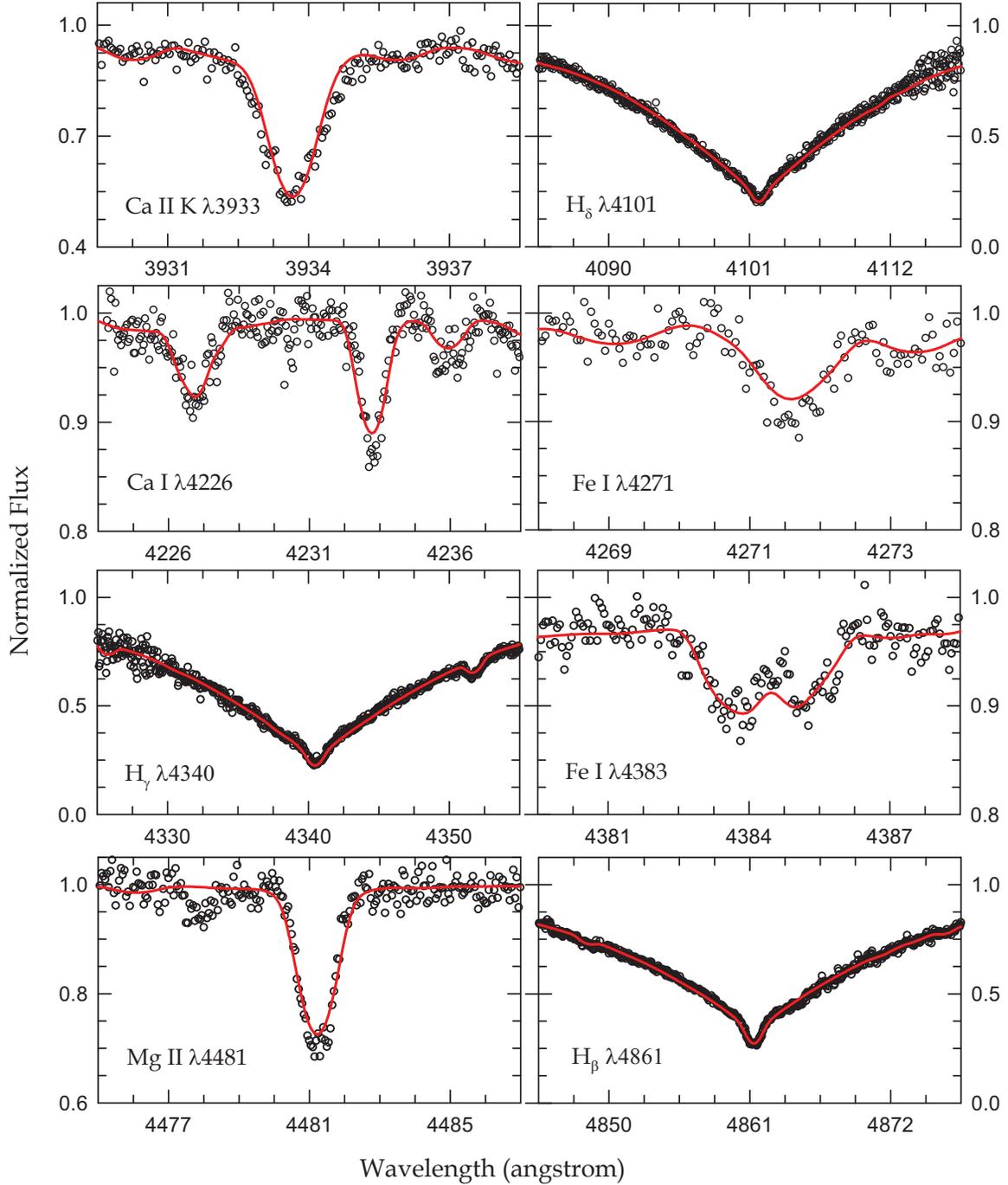}
\caption{Eight spectral regions of the primary star. The open circles are the disentangling spectrum obtained with the FDBinary code. 
The red lines represent the synthetic model spectrum of our best-fitting parameters $T_{\rm eff,1}$ = 10,000 K and $v_1\sin$$i$ = 55 km s$^{-1}$. }
\label{Fig5}
\end{figure}

\begin{figure}
\includegraphics{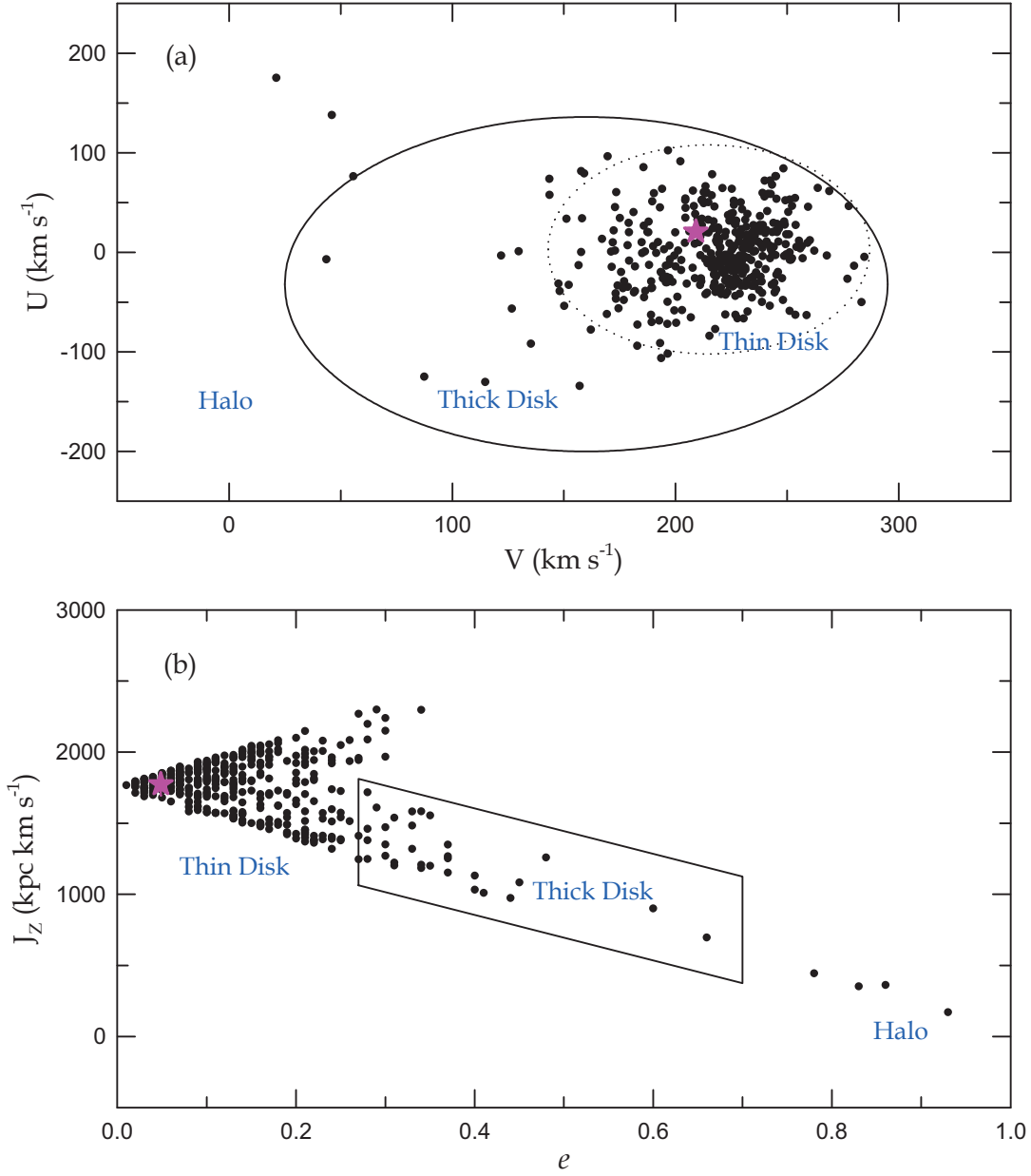}
\caption{(a) $U-V$ and (b) $J_z-e$ diagrams for WASP 0131+28 (star symbols) and 398 DA WDs (circles; Pauli et al. 2006). The dotted 
and solid eclipses in the upper panel (a) represent the 3$\sigma$ thin- and thick-disk contours, respectively, while the solid box in 
the lower panel (b) marks the thick-disk region (Pauli et al. 2006). }
\label{Fig6}
\end{figure}

\begin{figure}
\includegraphics{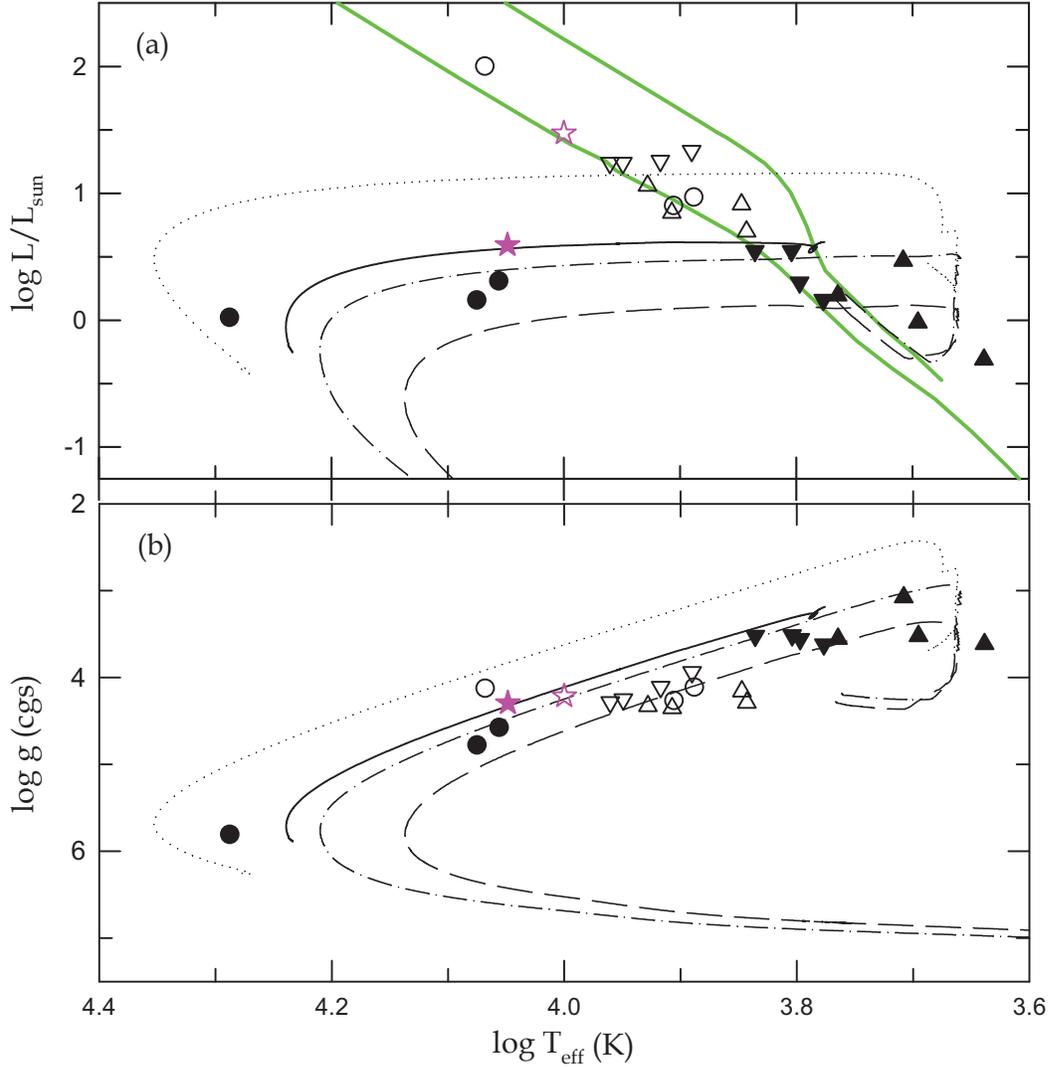}
\caption{(a) HR and (b) $\log T_{\rm eff}-\log g$ diagrams for WASP 0131+28 (star symbols), semi-detached and detached 
R CMa types (up- and down-pointing triangles), and EL CVn types (circles). The open and filled symbols represent the main-sequence 
dwarfs and the pre-He WDs, respectively. The dashed, dashed-dot, dotted, and solid lines are the evolutionary tracks of He WDs with 
masses of 0.179 M$_\odot$, 0.195 M$_\odot$, 0.234 M$_\odot$ (Driebe et al. 1998), and 0.203 M$_\odot$ (Althaus et al. 2013), 
respectively. In the upper panel (a), the solid green lines denote the ZAMS and TAMS for solar metallicity. }
\label{Fig7}
\end{figure}

\clearpage
\begin{deluxetable}{lcclc}
\tablewidth{0pt} 
\tablecaption{CCD Photometric Data of WASP 0131+28 Observed at LOAO }
\tablehead{
\colhead{BJD} & \colhead{$B$} && \colhead{BJD} & \colhead{$V$}
}
\startdata
2,457,710.55778  & 10.950   &&   2,457,710.55807  & 10.825       \\
2,457,710.55952  & 10.888   &&   2,457,710.55981  & 10.825       \\
2,457,710.56123  & 10.932   &&   2,457,710.56152  & 10.855       \\
2,457,710.56295  & 10.942   &&   2,457,710.56324  & 10.870       \\
2,457,710.56467  & 10.965   &&   2,457,710.56496  & 10.853       \\
2,457,710.56638  & 10.949   &&   2,457,710.56667  & 10.830       \\
2,457,710.56809  & 10.953   &&   2,457,710.56838  & 10.837       \\
2,457,710.57040  & 10.949   &&   2,457,710.57069  & 10.887       \\
2,457,710.57211  & 11.005   &&   2,457,710.57354  & 10.865       \\
2,457,710.57441  & 10.981   &&   2,457,710.58042  & 10.887       \\
\enddata
\tablecomments{This table is available in its entirety in machine-readable and Virtual Observatory (VO) forms.}
\end{deluxetable}

\begin{deluxetable}{lcc}
\tablewidth{0pt} 
\tablecaption{Standard Magnitudes and Color Indices of WASP 0131+28 }
\tablehead{
\colhead{Phase}  & \colhead{$V$ (mag)}  & \colhead{$B-V$ (mag)}  
}                                                                                                                                     
\startdata                                                                                                                            
0.00             & 10.929$\pm$0.015       &  0.092$\pm$0.023    \\
0.25             & 10.820$\pm$0.016       &  0.085$\pm$0.023    \\
0.50             & 10.905$\pm$0.017       &  0.087$\pm$0.025    \\
0.75             & 10.816$\pm$0.015       &  0.087$\pm$0.023    \\
\enddata
\end{deluxetable}

\begin{deluxetable}{lrcrc}
\tablewidth{0pt}                    
\tabletypesize{\small}   
\tablecaption{Radial Velocities of WASP 0131+28 }                                                                            
\tablehead{    
\colhead{BJD}          & \colhead{$V_{1}$}       & \colhead{$\sigma_1$}    & \colhead{$V_{2}$}       & \colhead{$\sigma_2$}    \\                                            
\colhead{(2,457,000+)} & \colhead{(km s$^{-1}$)} & \colhead{(km s$^{-1}$)} & \colhead{(km s$^{-1}$)} & \colhead{(km s$^{-1}$)}  
}                                                                                                 
\startdata
 408.9382             &  $ 11.26$                &  0.50                   &  $-212.25$              &  1.78                   \\
 408.9662             &  $ 12.01$                &  0.55                   &  $-208.30$              &  1.96                   \\
 409.0013             &  $ 10.99$                &  0.51                   &  $-202.76$              &  1.81                   \\
 409.0363             &  $  8.97$                &  0.77                   &  $-198.36$              &  2.69                   \\
 410.9235             &  $  8.66$                &  0.66                   &  $-194.92$              &  2.13                   \\
 410.9585             &  $  9.29$                &  0.52                   &  $-185.20$              &  1.67                   \\
 762.9656             &  $ 10.31$                &  0.37                   &  $-203.13$              &  1.31                   \\
 762.9937             &  $ 10.11$                &  0.35                   &  $-193.62$              &  1.30                   \\
 763.0695             &  $  6.82$                &  0.35                   &  $-169.18$              &  1.21                   \\
 763.9563             &  $-27.25$                &  0.31                   &  $ 169.82$              &  1.08                   \\
 763.9843             &  $-24.89$                &  0.32                   &  $ 162.58$              &  1.13                   \\
 764.0124             &  $-24.38$                &  0.34                   &  $ 151.61$              &  0.99                   \\
 764.0404             &  $-21.94$                &  0.39                   &  $ 137.35$              &  1.37                   \\
 765.0135             &  $  4.46$                &  0.36                   &  $-132.95$              &  1.21                   \\
 765.0415             &  $  3.54$                &  0.56                   &  $-125.41$              &  1.99                   \\
 804.9383             &  $-19.01$                &  0.53                   &  $ 107.64$              &  1.84                   \\
 808.9424             &  $-27.67$                &  0.41                   &  $ 192.40$              &  1.42                   \\
 808.9704             &  $-27.92$                &  0.41                   &  $ 193.31$              &  1.42                   \\
 809.9366             &  $ 12.98$                &  0.32                   &  $-208.39$              &  1.13                   \\
 809.9647             &  $ 12.14$                &  0.38                   &  $-207.26$              &  1.33                   \\
1017.0410             &  $ 13.85$                &  0.39                   &  $-208.36$              &  1.35                   \\
1017.0690             &  $ 12.57$                &  0.38                   &  $-207.21$              &  1.33                   \\
1017.0971             &  $ 11.80$                &  0.32                   &  $-206.01$              &  1.11                   \\
1017.1251             &  $ 10.43$                &  0.32                   &  $-199.97$              &  1.14                   \\
1078.9175             &  $  3.20$                &  0.39                   &  $-144.71$              &  1.39                   \\
1078.9455             &  $  6.56$                &  0.33                   &  $-155.05$              &  1.16                   \\
1078.9736             &  $  7.35$                &  0.36                   &  $-170.40$              &  1.27                   \\
1079.0016             &  $  8.72$                &  0.40                   &  $-182.86$              &  1.41                   \\
1079.0297             &  $ 10.33$                &  0.37                   &  $-189.61$              &  1.29                   \\
1079.0577             &  $  9.76$                &  0.33                   &  $-198.04$              &  1.16                   \\
1079.0858             &  $ 12.49$                &  0.30                   &  $-202.00$              &  1.06                   \\
1079.1138             &  $ 11.83$                &  0.36                   &  $-205.23$              &  1.27                   \\
1079.1419             &  $ 12.73$                &  0.34                   &  $-212.35$              &  1.19                   \\
1079.1699             &  $ 11.43$                &  0.37                   &  $-207.01$              &  1.33                   \\
1783.1039             &  $  6.68$                &  0.36                   &  $-157.78$              &  1.28                   \\
1783.1319             &  $  6.51$                &  0.34                   &  $-165.63$              &  1.22                   \\
1783.1602             &  $  8.47$                &  0.46                   &  $-177.20$              &  1.62                   \\
1783.1883             &  $  9.71$                &  0.39                   &  $-193.78$              &  1.27                   \\
1783.2165             &  $  9.48$                &  0.36                   &  $-197.12$              &  1.26                   \\
1783.2447             &  $ 10.81$                &  0.47                   &  $-202.56$              &  1.65                   \\
1784.0444             &  $-23.88$                &  0.27                   &  $ 140.34$              &  0.96                   \\
1784.0725             &  $-25.16$                &  0.26                   &  $ 156.40$              &  0.92                   \\
1784.1006             &  $-25.89$                &  0.25                   &  $ 164.46$              &  0.89                   \\
1784.1287             &  $-26.72$                &  0.30                   &  $ 175.61$              &  1.01                   \\
1784.1568             &  $-27.81$                &  0.22                   &  $ 181.47$              &  0.78                   \\
1784.1849             &  $-28.42$                &  0.28                   &  $ 184.72$              &  0.99                   \\
1784.2130             &  $-27.89$                &  0.39                   &  $ 190.17$              &  1.37                   \\
1784.2411             &  $-29.09$                &  0.24                   &  $ 192.73$              &  0.86                   \\
1784.2692             &  $-29.36$                &  0.36                   &  $ 193.37$              &  1.25                   \\
1784.2973             &  $-29.04$                &  0.27                   &  $ 190.82$              &  0.94                   \\
\enddata                                                                                                             
\end{deluxetable}

\begin{deluxetable}{lcc}
\tablewidth{0pt} 
\tablecaption{Light and RV Parameters of WASP 0131+28 }
\tablehead{
\colhead{Parameter}               & \colhead{Primary}  & \colhead{Secondary}                                                  
}                                                                                                                                     
\startdata                                                                                                                            
$T_0$ (BJD)                       & \multicolumn{2}{c}{2,457,711.56158$\pm$0.00083}   \\
$P$ (day)                         & \multicolumn{2}{c}{1.8827597$\pm$0.0000015}       \\
$a$ (R$_\odot$)                   & \multicolumn{2}{c}{8.324$\pm$0.019}               \\
$\gamma$ (km s$^{-1}$)            & \multicolumn{2}{c}{$-$8.44$\pm$0.15}              \\
$K_1$ (km s$^{-1}$)               & \multicolumn{2}{c}{20.36$\pm$0.20}                \\
$K_2$ (km s$^{-1}$)               & \multicolumn{2}{c}{202.57$\pm$0.60}               \\
$q$                               & \multicolumn{2}{c}{0.1005$\pm$0.0010}             \\
$i$ (deg)                         & \multicolumn{2}{c}{85.027$\pm$0.020}              \\
$T$ (K)                           & 10,000$\pm$200     & 11,186$\pm$235               \\
$\Omega$                          & 4.6783$\pm$0.0034  & 3.0323$\pm$0.0027            \\
$\Omega_{\rm in}$$\rm ^a$         & \multicolumn{2}{c}{1.9606}                        \\
$A$                               & 1.0                & 1.0                          \\
$g$                               & 1.0                & 1.0                          \\
$X_{\rm bol}$, $Y_{\rm bol}$      & 0.666, 0.075       & 0.700, 0.068                 \\
$x_{B}$, $y_{B}$                  & 0.705, 0.335       & 0.661, 0.338                 \\
$x_{V}$, $y_{V}$                  & 0.605, 0.290       & 0.567, 0.287                 \\
$x_{T_{\rm P}}$, $y_{T_{\rm P}}$  & 0.409, 0.198       & 0.387, 0.193                 \\
$L/(L_1+L_2)_{B}$                 & 0.9009$\pm$0.0010  & 0.0991                       \\
$L/(L_1+L_2)_{V}$                 & 0.9048$\pm$0.0008  & 0.0952                       \\
$L/(L_1+L_2)_{T_{\rm P}}$         & 0.9079$\pm$0.0002  & 0.0921                       \\
$r$ (pole)                        & 0.2183$\pm$0.0002  & 0.0634$\pm$0.0001            \\
$r$ (point)                       & 0.2200$\pm$0.0002  & 0.0637$\pm$0.0001            \\
$r$ (side)                        & 0.2196$\pm$0.0002  & 0.0634$\pm$0.0001            \\
$r$ (back)                        & 0.2199$\pm$0.0002  & 0.0637$\pm$0.0001            \\
$r$ (volume)$\rm ^b$              & 0.2193$\pm$0.0002  & 0.0635$\pm$0.0001            \\ 
\enddata
\tablenotetext{a}{Potential for the inner critical Roche surface.}
\tablenotetext{b}{Mean volume radius.}
\end{deluxetable}

\begin{deluxetable}{lcc}
\tablewidth{0pt} 
\tablecaption{Absolute Parameters of WASP 0131+28 }
\tablehead{
\colhead{Parameter}           & \colhead{Primary}   & \colhead{Secondary}                                                  
}                                                                                                                                     
\startdata                                                                                                                            
$M$ ($M_\odot$)               & 1.986$\pm$0.017     & 0.200$\pm$0.002             \\
$R$ ($R_\odot$)               & 1.824$\pm$0.006     & 0.528$\pm$0.003             \\
$\log$ $g$ (cgs)              & 4.214$\pm$0.004     & 4.293$\pm$0.007             \\
$\rho$ ($\rho_\odot$)         & 0.328$\pm$0.004     & 1.357$\pm$0.027             \\
$L$ ($L_\odot$)               & 29.8$\pm$2.4        & 3.91$\pm$0.33               \\
$M_{\rm bol}$ (mag)           & 1.04$\pm$0.09       & 3.25$\pm$0.10               \\
BC (mag)                      & $-$0.25$\pm$0.04    & $-$0.51$\pm$0.05            \\
$M_{\rm V}$ (mag)             & 1.29$\pm$0.10       & 3.76$\pm$0.11               \\
Distance (pc)                 & \multicolumn{2}{c}{766$\pm$35}                    \\
\enddata
\end{deluxetable}

\begin{deluxetable}{lcccccccc}
\tablewidth{0pt}
\tabletypesize{\small}
\tablecaption{Kinematical Data for WASP 0131+28 }
\tablehead{
\colhead{$X$}   & \colhead{$Y$} & \colhead{$Z$}    & \colhead{$U$}  & \colhead{$V$} & \colhead{$W$}   & \colhead{$J_z$}   & \colhead{$e$} \\
\colhead{(kpc)} & (kpc)         & (kpc)            & (km s$^{-1}$)  & (km s$^{-1}$) & (km s$^{-1}$)   & (kpc km s$^{-1}$) &               
}
\startdata
$-$8.4$\pm$0.4  & 0.46$\pm$0.02 & $-$0.43$\pm$0.02 & 20.8$\pm$0.4   & 209.1$\pm$0.8 & $-$12.4$\pm$1.2 & 1774$\pm$79       & 0.049$\pm$0.002  \\
\enddata
\tablecomments{Right-handed coordinate system. $U$, $V$, and $W$ are positive in the directions of the Galactic center, Galactic rotation, and the North Galactic Pole, respectively. }
\end{deluxetable}


\newpage
\begin{thebibliography}{}
\bibitem[Althaus et al(2013)]{althaus2013} Althaus, L. G., Miller Bertolami, M. M., \& C\'orsico, A. H. 2013, A\&A, 557, A19
\bibitem[Bohlin et al(2017)]{bohlin2017} Bohlin, R. C., M{\'e}sz{\'a}ros, S., Fleming, S. W., et al.\ 2017, AJ, 153, 234 
\bibitem[Bradley et al(2017)]{bradley2017} Bradley, L., Sipocz, B., Robitaille, T., et al. 2017, astropy/photutils: v0.4, Zenodo, doi:10.5281/zenodo.1039309
\bibitem[Budding \& Butland(2011)]{budding2011} Budding, E., \& Butland, R. 2011, MNRAS, 418, 1764
\bibitem[Castelli \& Kurucl(2003)]{castelli2003} Castelli, F., \& Kurucz, R. L. 2003, in IAU Symp. 210, Modelling of Stellar Atmospheres, ed. N. Piskunov, W. W. Weiss, \& D. F. Gray (Uppsala: Uppsala Univ.), 20
\bibitem[Chen et al(2017)]{chen2017} Chen, X., Maxted, P. F. L., Li, J., \& Han, Z. 2017, MNRAS, 467, 1874
\bibitem[Driebe et al(1998)]{driebe1998} Driebe, T., Sch\"oenberner, D., Bl\"oecker, T., \& Herwig, F. 1998, A\&A, 339, 123
\bibitem[Fontaine et al(2001)]{fontaine2001} Fontaine, G., Brassard, P., \& Bergeron, P. 2001, PASP, 113, 409
\bibitem[GAIA(2018)]{gaia2018} Gaia Collaboration, Brown, A. G. A., Vallenari, A., et al. 2018, A\&A, 616, A1
\bibitem[Gilliland \& Brown(1988)]{gilliland1988} Gilliland, R. L., \& Brown, T. M. 1988, PASP, 100, 754 
\bibitem[Guo et al(2017)]{guo2017} Guo, Z., Gies, D. R., Matson, R. A., Garcia Hernandez, A., Han, Z., \& Chen, X. 2017, ApJ, 837, 114
\bibitem[Henden et al(2016)]{henden2016} Henden, A. A., Templeton, M., Terrell, D., et al. 2016, yCat, 2336, 0 
\bibitem[Hilditch(2001)]{hilditch2001} Hilditch, R. W. 2001, An Introduction to Close Binary Stars (Cambridge: Cambridge Univ. Press)
\bibitem[Hong et al(2017)]{hong2017} Hong, K., Lee, J. W., Koo, J.-R., et al. 2017, AJ, 153, 247
\bibitem[Ilijic et al(2004)]{ilijic2004} Iliji\'c, S., Hensberge, H., Pavlovski, K., \& Freyhammer, L. M. 2004, in ASP Conf. Ser. 318, Spectroscopically and Spatially Resolving the Components of the Close Binary Stars, ed. R. Hilditch, H. Hensberge, \& K. Pavlovski (San Francisco, CA: ASP), 111 
\bibitem[Johnson \& Soderblom(1987)]{johnson1987} Johnson, D. R. H., \& Soderblom, D. R. 1987, AJ, 93, 864
\bibitem[Kepler et al(2007)]{kepler2007} Kepler, S. O., Kleinman, S. J., Nitta, A., et al. 2007, MNRAS, 375, 1315 
\bibitem[Kilic et al(2007)]{kilic2007} Kilic, M., Stanek, K. Z., \& Pinsonneault, M. H. 2007, ApJ, 671, 761
\bibitem[Kim et al(2007)]{kim2007} Kim, K.-M., Han, I., Valyavin, G. G., et al. 2007, PASP, 119, 1052
\bibitem[Koo et al(2014)]{koo2014} Koo, J.-R., Lee, J. W., Lee, B.-C., et al. 2014, AJ, 147, 104
\bibitem[Lee \& Park(2018)]{lee2018} Lee, J. W., \& Park, J.-H. 2018, MNRAS, 480, 4693
\bibitem[Lee et al(2017)]{lee2017} Lee, J. W., Hong, K., Kim, S.-L., \& Koo, J.-R. 2017, ApJ, 835, 189
\bibitem[Lee et al(2018)]{lee2018} Lee, J. W., Hong, K., Koo, J.-R., \& Park, J.-H. 2018, AJ, 155, 5
\bibitem[Lee et al(2016)]{lee2016} Lee, J. W., Kim, S.-L., Hong, K., Koo, J.-R., Lee, C.-U., \& Youn, J.-H. 2016, AJ, 151, 25
\bibitem[Lee et al(2019)]{lee2019} Lee, J. W., Kristiansen, M., \& Hong, K. 2019, AJ, 157, 223
\bibitem[Lee et al(2012)]{lee2012} Lee, J. W., Youn, J.-H., Kim, S.-L., Lee, C.-U., \& Hinse, T. C. 2012, AJ, 143, 95 
\bibitem[Lehmann et al(2013)]{lehmann2013} Lehmann, H., Southworth, J., Tkachenko, A., \& Pavlovski, K. 2013, A\&A, 557, A79
\bibitem[Li et al(2019)]{li2019} Li, Z., Chen, X., Chen, H.-L., \& Han, Z. 2019, ApJ, 871, 148
\bibitem[Lin et al(2011)]{lin2011} Lin, J., Rappaport, S., Podsiadlowski, P., et al. 2011, ApJ, 732, 70
\bibitem[Marsh et al(1995)]{marsh1995} Marsh, T. R., Dhillon, V. S., \& Duck, S. R. 1995, MNRAS, 275, 828
\bibitem[Matson et al(2015)]{matson2015} Matson, R. A., Gies, D. R., Guo, Z., et al. 2015, ApJ, 806, 155
\bibitem[Maxted et al(2011)]{maxted2011} Maxted, P. F. L., Anderson, D. R., Burleigh, M. R., et al. 2011, MNRAS, 418, 1156
\bibitem[Maxted et al(2014a)]{maxted2014a} Maxted, P. F. L., Bloemen, S., Heber, U., et al. 2014a, MNRAS, 437, 168
\bibitem[Maxted et al(2014b)]{maxted2014b} Maxted, P. F. L., Serenelli, A. M., Marsh, T. R., Catal\'an, S., Mahtani, D. P., \& Dhillon, V. S. 2014b, MNRAS, 444, 208
\bibitem[Maxted et al(2013)]{maxted2013} Maxted, P. F. L., Serenelli, A. M., Miglio, A., et al. 2013, Natur, 498, 463
\bibitem[Mochnacki(1984)]{mochnacki1984} Mochnacki, S. W. 1984, ApJS, 55, 551
\bibitem[Odenkirchen \& Brosche(1992)]{odenkirchen1992} Odenkirchen, M., \& Brosche, P. 1992, AN, 313, 69
\bibitem[Pauli et al(2006)]{pauli2006} Pauli, E.-M., Napiwotzki, R., Heber, U., Altmann, M., \& Odenkirchen, M. 2006, A\&A, 447, 173
\bibitem[Reid(1993)]{reid1993} Reid, M. J. 1993, ARA\&A, 31, 345
\bibitem[Ricker et al(2015)]{ricker2015} Ricker, G. R., Winn, J. N., Vanderspek, R., et al. 2015, JATIS, 1, 014003
\bibitem[Schlafly \& Finkbeiner(2011)]{schlafly2011} Schlafly, E. F., \& Finkbeiner, D. P. 2011, AJ, 737, 103
\bibitem[Southworth (2010)]{southworth2010} Southworth, J. 2010, MNRAS, 408, 1689
\bibitem[Stassun et al(2018)]{stassun2018} Stassun, K. G., Oelkers, R. J., Pepper, J., et al. 2018, AJ, 156, 102
\bibitem[Stassun et al(2019)]{stassun2019} Stassun, K. G., Oelkers, R. J., Paegert, M., et al. 2019, AJ, 158, 138
\bibitem[Temmink et al(2019)]{temmink2019} Temmink, K. D., Toonen, S., Zapartas, E., et al. 2019, arXiv:1910.05335 
\bibitem[Tian et al(2015)]{tian2015} Tian, H.-J., Liu, C., Carlin, J. L., et al. 2015, ApJ, 809, 145
\bibitem[Torres(2010)]{torres2010} Torres, G. 2010, AJ, 140, 1158
\bibitem[Torres et al(2010)]{torres2010} Torres, G., Andersen, J., \& Gim\'enez, A. 2010, A\&ARv, 18, 67
\bibitem[van Dokkum(2001)]{van2001} van Dokkum, P. G. 2001, PASP, 113, 1420 
\bibitem[van Hamme(1993)]{vanHamme1993} van Hamme, W. 1993, AJ, 106, 209
\bibitem[van Hamme \& Wilson(2007)]{vanHamme2007} van Hamme, W., \& Wilson, R. E. 2007, ApJ, 661, 1129
\bibitem[van Kerkwijk et al(2010)]{vanKerkwijk2010} van Kerkwijk, M. H., Rappaport, S. A., Breton, R. P., et al. 2010, ApJ, 715, 51
\bibitem[van Roestel et al(2018)]{vanRoestel2018} van Roestel, J., Kupfer, T., Ruiz-Carmona, R., et al. 2018, MNRAS, 475, 2560
\bibitem[Wang et al(2020)]{wang2020} Wang, K., Zhang, X., \& Dai, M. 2020, ApJ, 888, 49
\bibitem[Wang et al(2019)]{wang2019} Wang, K., Zhang, X., Luo, Y., \& Luo, C. 2019, MNRAS, 486, 2462
\bibitem[Wilson \& Devinney(1971)]{wilson1971} Wilson, R. E., \& Devinney, E. J. 1971, ApJ, 166, 605
\bibitem[Woosley \& Heger(2015)]{woosley2015} Woosley, S. E., \& Heger, A. 2015, ApJ, 810, 34
\bibitem[Zhang et al(2016)]{Zhang2016} Zhang, X. B., Fu, J. N., Li, Y., Ren, A. B., \& Luo, C. Q. 2016, ApJL, 821, L32
\bibitem[Zucker \& Mazeh(1994)]{Zucker1994} Zucker, S., \& Mazeh, T. 1994, ApJ, 420, 806
\end{thebibliography}
\end{document}